\documentstyle[aps,prl,multicol]{revtex}
\include{psfig}
\input epsf
\begin{document}
\author{G. Foffi$^1$, K.A. Dawson$^1$, 
S.V. Buldyrev$^{2,3}$, F. Sciortino$^2$, 
E. Zaccarelli$^2$, P. Tartaglia$^2$}
\address{$^1$ Irish Centre for Colloid Science
and Biomaterials, Department of Chemistry,\\ University College Dublin,
Belfield, Dublin 4, Ireland\\
$^2$ Dipartimento di Fisica, Istituto Nazionale di Fisica della Materia,
and INFM Center for Statistical Mechanics \\ and Complexity,
Universit\`{a} di Roma 
La Sapienza, P.le A. Moro 5, I-00185 Rome, Italy\\
$^3$ Center for Polymer Studies and Department of Physics, Boston University,
Boston, MA 02215, USA.
}
\title{{\bf  Evidence for Unusual Dynamical Arrest Scenario in 
Short Ranged Colloidal Systems}
}
\date{\today}
\maketitle

\begin{abstract}
Extensive molecular dynamics simulation studies of 
particles interacting via a short ranged
attractive square-well (SW) potential are reported. 
The calculated loci of constant diffusion coefficient  $D$ 
in the temperature-packing fraction
plane show a re-entrant behavior , i.e. an increase of diffusivity on
cooling, confirming an important part of the high
volume-fraction dynamical-arrest scenario earlier predicted by
theory for particles with short ranged potentials.
The more efficient localization 
mechanism induced by the short range bonding provides, on average, 
additional free volume as compared to the hard-sphere case and results
in faster dynamics.
\end{abstract}


64.70.Pf Glass transitions
82.70.Dd Colloids.

\begin{multicols}{2}
\smallskip%
Recently a large number of novel dynamical arrest phenomena have been 
described in
systems where, besides the usual hard core, an attractive potential is
present with a range much smaller than the hard core radius
\cite{fabbian99,bergenholtz99,foffi01}. 
This condition is not usually met in molecular liquid
systems, where the attractive range is of the same order of magnitude as 
the hard core, but
can be realized in colloidal systems where the size of the particles 
largely exceeds the range of the attractive part of the 
potential\cite{lekkerkerker,poon,bartschlog,bartsch,verduin95}, and possibly
many other systems of experimental interest, including globular
proteins \cite{proteincrystallization,buckyballs}.
%
%
From the
experimental point of view the most striking phenomenon associated
with the attraction is the gradual disappearance of the liquid phase
when the range of the attractive part of the potential diminishes. 
Ultimately, for narrower wells, the liquid-gas coexistence becomes
metastable with respect to a crystal-fluid equilibrium, but
nevertheless shows up as a metastable binodal curve in
experiments \cite{verduin95} and in simulations \cite{auer01}.
Various approximations to the liquid and crystalline free
energies have been used in order to calculate the coexistence lines
and for various forms of the attractive part of the interaction
potential\cite{barker,gast,evans,foffi02}.  

%
%
While the situation of the equilibrium phase diagram is quite
clear, the metastable region of the phase-diagram related to the supercooled
fluid is quite complex, and only partially clarified 
\cite{verduin95,segre,lai}.
It is now widely accepted that mode-coupling theory (MCT) provides
a  description of supercooled dynamics which is particularly
suitable for understanding colloidal systems. MCT predicts in
particular the existence of a non-ergodicity transition which
corresponds to a structural arrest, i.e. the impossibility of the
particles of the system to move under the effect of the
neighboring ones, the so-called cage
effect\cite{goetze91,goetze92}. Some measurements performed in
colloidal systems support the MCT predictions in more quantitative
detail\cite{goetze99,vanmengen9394}.
%
%
More recently MCT has been applied to systems interacting through
a short ranged attractive potential, with the result of predicting a number of
new and interesting phenomena
\cite{fabbian99,bergenholtz99,dawson00,fuchscondmat}. Amongst the
most striking is the possibility to distinguish two
types of transition from a supercooled liquid to a glass, one mainly
due to the repulsive part of the potential through the usual mechanism 
of the cage effect, the other  to the attractive part of the potential.
The latter mechanism is due to the adhesiveness of the potential 
at short distances which produces `clusters' 
of particles, a different mechanism of structural arrest than 
the `blocking' or `jamming' familiar in hard sphere
systems. The attractive branch of glass transition curve that
results extends from low values of the volume fraction $\phi$ to
values of the order of the hard sphere transition and does not
vary much with temperature. The repulsive glass curve passes from
the hard sphere value at high temperatures to larger values of
$\phi$ as temperature decreases, thereby giving rise to a
re-entrant phenomenon, i.e. the supercooled liquid phase extends
into the glass region above the volume fractions of the pure hard
spheres system. It is then possible, raising the temperature at
constant volume fraction, to move from the attractive glass region
to the supercooled fluid one and again in the amorphous
hard-sphere-like glass region.  We note in passing that the
complementary possibility of driving the system across a glass
transition both by increasing and decreasing the density has been
shown to be possible in systems with long range interactions
(Wigner glasses)\cite{bosse}.
%
%
In the case here under scrutiny, at high values of $\phi$, the two
branches of the glass curves cross at an angle, and the attractive
branch continues further into the glass region, giving rise to a
remarkable coexistence of two different types of glass which can
be characterized, for example, by their mechanical properties such
as the shear modulus\cite{zaccarelli01}. The glass-glass curve
terminates in a special singular point of the theory, named $A_3$
in the MCT, where the relaxation processes have a peculiar
behavior \cite{sjoegren}.
Upon narrowing the width of the attractive
potential, the glass-glass transition line tends to become shorter
until it vanishes in a point which represents a high order 
singularity, an $A_4$ singularity in the MCT language \cite{dawson00}.
%
%
The experimental and/or numerical verifications of these predictions
are still scarce\cite{fuchscondmat,bartschlog}. A logarithmic decay,
predicted by MCT close to the $A_3$ point, was detected in particular
in a micellar system at high packing fractions\cite{mallamace00}. It
is interesting to note the possible application of this type of
results to the study of protein crystallization
\cite{foffi02,proteincrystallization}.

We focus in this paper on a particular aspect of the attractive
colloidal system, that is the extension of the glass transition line 
to high values of the packing fraction, in order to test one of the important 
predictions of the theory, i.e. the re-entrant behavior of the 
supercooled fluid-glass line.
%
%
We simulate a monodisperse sample of $N=1237$ particles of unit
mass with a constant diameter $\sigma=1$ in a cubic box. 
The physical quantities are measured in units of the particle diameter 
$\sigma$, the particle mass $m$, and the square well depth $u_0$ 
as unit of energy. Temperature $T$ is measured in units of energy,
i.e. by setting the Boltzmann constant $k_B=1$.
With these choices, time is measured in units of 
$\sigma (m/u_o)^{1/2}$. The interparticle potential $V(r)$ is
the square well potential,
\begin{eqnarray}
V(r)&=&\infty~~~~~~~~~~~~~~~~~~ r<\sigma \nonumber\\
V(r)&=&-u_0~~~~~~\sigma<r<\sigma+\Delta \nonumber\\
V(r)&=&0~~~~~~~~~~~~~r>\sigma+\Delta
\end{eqnarray}

The width of the attractive part of the potential is $\Delta=0.0309$, 
corresponding to 
a percentage variation $\epsilon=\Delta/(\sigma+\Delta)=0.03$.  
We have investigated volume fractions from $0.10$ to $\approx0.58$ and 
temperatures from $T=0.32$ to $50$. For $T$ lower than $0.32$ the homogeneous
fluid phase is unstable with respect to gas-liquid
phase-separation. 
MCT calculations, based both on the Percus-Yevick (PY) and Mean Spherical 
Approximation (MSA) structure factors \cite{dawson00}, 
predict, for this specific potential, a reentrant fluid-glass line, 
as discussed in what follows.
%
%
We have implemented the standard MD algorithm for particles
interacting with SW potentials\cite{rapaport}. Between collisions,
particles move along straight lines with constant velocities. When the
distance between the particles becomes equal to the distance for which
$V(r)$ has a discontinuity, the velocities of the interacting
particles instantaneously change.  The algorithm calculates the
shortest collision time in the system and propagate the trajectory
from one collision to the next one.  Calculations of the next
collision time are optimized by dividing the system in small
subsystems, so that collision time are computed only between particles
in the neighboring subsystems.

For each run we have used particular care in equilibrating the system,
and starting from an equilibrated configuration we have performed a 
simulation up to a time  $10^3$. 
Runs exhibiting the
presence of crystalline nucleus at the end of the simulations were
discarded. Since we simulate a monodisperse system, 
the formation of a crystal phase fixes the range of packing fractions
where a stable (or metastable) fluid phase can be studied.

We have calculated the self-diffusion coefficient $D$ in the 
supercooled liquid phase via the long time limit of 
the mean squared displacement of the particles. 
For each of the 10 studied isotherms,  
$D$ varies almost over three decades, showing a marked
decrease at high volume fractions, before crystallization takes place.

\begin{figure}
\centerline{\psfig{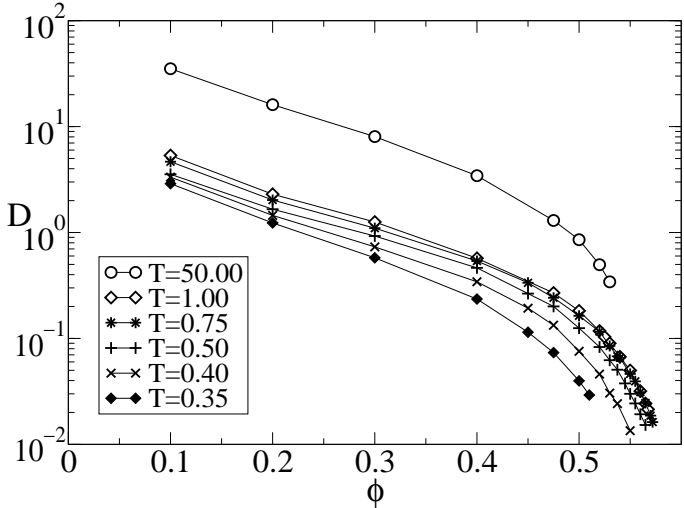}
           }  
\caption{
Diffusion coefficient $D$ as a function of the 
packing fraction $\phi$ for the $\epsilon=0.03$ square well
potential at several different $T$. 
}
\label{fig:D}
\end{figure}

Fig.\ref{fig:D} shows the diffusion coefficient $D$ as a function of the
packing fraction $\phi \equiv N \pi \sigma^3/(6V)$ for the
studied isotherms. The $T=50$ isotherm diffusivity reproduces the
hard sphere behavior. For each isotherm, 
simulations at larger volume fraction than the ones reported 
inevitably lead to crystallization during the run.
Fig.\ref{fig:D} shows that, when the kinetic energy is of the
order of the potential depth (e.g. $T=0.75$) , 
crystallization is shifted to larger
packing fraction values as compared to both the hard-sphere case 
(e.g. $T=50$) and to low $T$ (e.g. $T=0.35$).

\begin{figure}
\centerline{\psfig{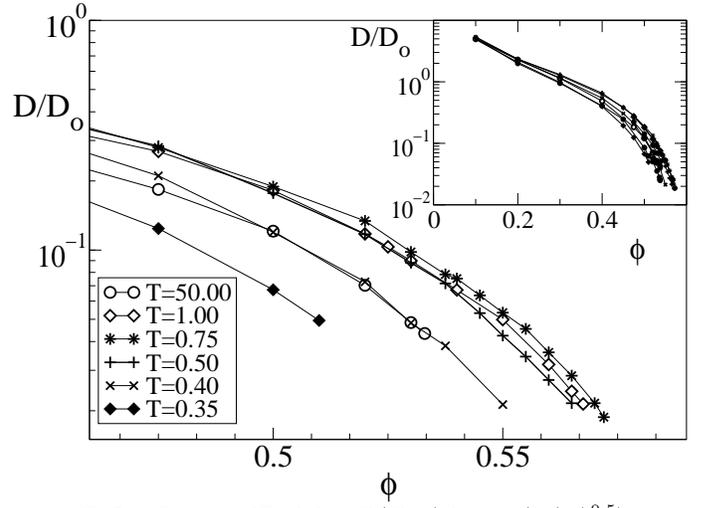}
           }  
\caption{
Scaled diffusivity $D/D_o$  $(D_o=\sigma (T/m)^{0.5})$ 
as a function of packing fraction $\phi$ for the SW potential for some of the 
studied isotherms. 
The inset shows an enlarged $\phi$ window to highlight the
common low density limit.
}
\label{fig:2}
\end{figure}

Fig.\ref{fig:2} shows the diffusivity behavior in the $\phi$ region where the
reentrant phenomenon takes place. Data are
normalized  by $D_o\equiv \sigma \sqrt{T/m}$, 
in order to take into account the $T$ dependence of the microscopic time.

We see that on decreasing $T$ at constant $\phi$, $D/D_o$ first
increases and then decreases again.  Since the $T^{1/2}$ term in $D_o$
accounts already for the slowing down of the dynamics associated to
the different average particle velocity, the increase of $D/D_o$ on
cooling must have a different origin. This peculiar feature can be
explained as a result of the competition between two different
dynamical features produced by the increased bonding: (i) slowing down
of the dynamics due to the formation of a larger number of bonded
pairs and (ii) speeding up of the dynamics due to the larger free
volume resulting from the more efficient packing of the bonded
particles--- whose nearest neighbor distance is now imposed by the
short range of the potential.  The inset of Fig.\ref{fig:2} shows all
the studied state points.  At all $T$, the low density limit coincides
with the hard sphere behavior.



\begin{figure}
\centerline{\psfig{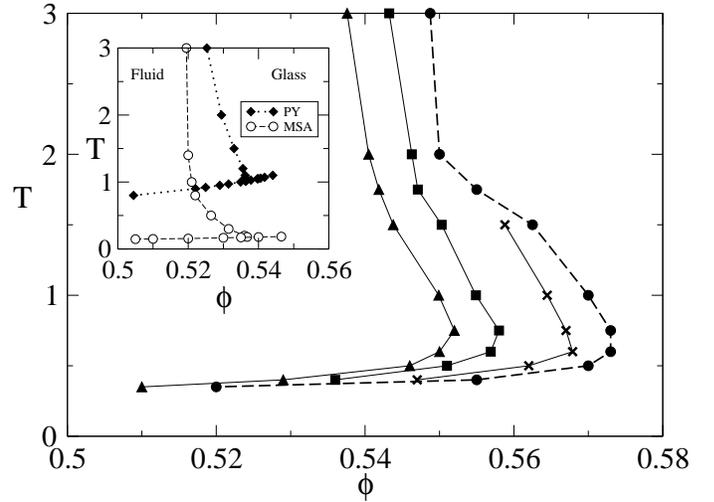}
           }  
\caption{
Iso diffusivity curves in the $(\phi,T)$ phase diagram with
$D/D_o=0.05$(triangles), $0.04$(squares), $0.025$(crosses). The dashed
line with filled circles represents the line where the system crystallizes
within our maximum simulation time.  The inset shows the theoretical
MCT prediction for the ideal glass line (redrawn from
Ref.\protect\cite{dawson00}) for both Percus Yevick (PY-filled
diamonds) and Mean Spherical Approximation (MSA-open circles),
separating the fluid phase from the glass phase.  }
\label{fig:3}
\end{figure}

Fig.~\ref{fig:3} reports the isodiffusivity curves, i.e. curves at
constant $D/D_0$ value, in the ($\phi,T$) plane.  It also reports the
curve where crystallization takes place within the time of our
calculation.  The isodiffusivity curves shown in Fig.~\ref{fig:3} can
be considered as the precursor of the fluid-glass transition line that
would take place when $D/D_0 \rightarrow 0$ if crystallization would
not occur first.  The inset of Fig.~\ref{fig:3} shows the MCT
calculations for the same model as reported in \cite{dawson00}. As in
all cases studied previously, the MCT calculation underestimates the
location of the glass transition line but provides a correct frame for
understanding the reentrant behavior observed in the present
calculations.

In summary, we have established the presence of reentrance
in the  dynamical arrest behavior of an extremely simple
model, the square well one, when the width of the
attractive potential is much smaller than the hard-core radius.
This condition can be met in colloidal systems. 
This observed behavior, predicted by MCT,  along with
a number of other associated phenomena, is explained as
competition between the hard-core caging --- characteristic of
the hard-core systems --- and the bonding caging which, in the case
of very short range potential, localizes the particles in a
more efficient manner than the hard core case.
This stronger localization, imposed by
incipient 'bonding' provides extra free volume which leads to more
diffusional pathways. Further lowering of the temperature produces
stronger cages, and the system then crosses over to the attractive
glass scenario. 
It is an open challenge to find out in future if 
other MCT predictions are supported by numerical investigation.
This additional work, which requires the study of larger values
of $\phi$ and larger simulation time windows to better characterize 
the slow dynamics --- based on SW
binary mixture systems to prevent crystallization ---
is underway.

%
This research is supported by the INFM-HOP-1999, MURST-PRIN-2000 and
COST P1. S.B. thanks the University of Rome for granting him a
visiting Professorship position.

\newpage

\vspace{10cm}

\vspace{3.5cm}

\end{multicols}
\end{document}